\documentclass[preprint,5p,twocolumn]{elsarticle}
\pdfoutput=1
\UseRawInputEncoding
\usepackage{geometry}
\usepackage{graphicx}
\usepackage{hyperref}
\usepackage{natbib}
\usepackage{bm}
\usepackage{orcidlink}
\usepackage{multirow}

\begin{document}

\begin{frontmatter}

\title{Deep Learning Detection and Classification of Gravitational Waves from \\
Neutron Star-Black Hole Mergers}

\author[1,2,3]{Richard Qiu \orcidlink{0000-0003-3462-0817}}
\ead{rqiu@college.harvard.edu}

\author[4,5]{Plamen G. Krastev \orcidlink{0000-0002-0830-1479}\corref{cor1}}
\ead{plamenkrastev@fas.harvard.edu}

\author[1,5]{Kiranjyot Gill \orcidlink{0000-0003-4341-9824}}
\ead{kiranjyot.gill@cfa.harvard.edu}

\author[1,5]{Edo Berger \orcidlink{0000-0002-9392-9681}}
\ead{eberger@cfa.harvard.edu}
\cortext[cor1]{Corresponding author}

\address[1]{Center for Astrophysics \textbar{} Harvard \& Smithsonian, 60 Garden Street, Cambridge, MA 02138-1516, USA}
\address[2]{Department of Physics, Harvard University, 17 Oxford Street Cambridge, MA 02138, USA}
\address[3]{John A. Paulson School of Engineering and Applied Sciences, Harvard University, 150 Western Ave, Allston, MA 02134, USA}
\address[4]{Faculty of Arts and Sciences Research Computing, Harvard University, 52 Oxford Street, Cambridge, MA 02138, USA}
\address[5]{The NSF AI Institute for Artificial Intelligence and Fundamental Interactions}

\begin{abstract}
The Laser Interferometer Gravitational-Wave Observatory (LIGO) and Virgo Interferometer Collaborations have now detected all three classes of compact binary mergers: binary black hole (BBH), binary neutron star (BNS), and neutron star-black hole (NSBH). For coalescences involving neutron stars, the simultaneous observation of gravitational and electromagnetic radiation produced by an event, has broader potential to enhance our understanding of these events, and also to probe the equation of state (EOS) of dense matter. However, electromagnetic follow-up to gravitational wave (GW) events requires rapid real-time detection and classification of GW signals, and conventional detection approaches are computationally prohibitive for the anticipated rate of detection of next-generation GW detectors. In this work, we present the first deep learning based results of classification of GW signals from NSBH mergers in \textit{real} LIGO data. We show for the first time that a deep neural network can successfully distinguish all three classes of compact binary mergers and separate them from detector noise. Specifically, we train a convolutional neural network (CNN) on $\sim 500,000$ data samples of real LIGO noise with injected BBH, BNS, and NSBH GW signals, and we show that our network has high sensitivity and accuracy. Most importantly, we successfully recover the two confirmed NSBH events to-date (GW191219 and GW200115) and the two confirmed BNS mergers to-date (GW170817 and GW190425), together with $\sim 90\%$ of all BBH candidate events from the third Gravitational Wave Transient Catalog, GWTC-3. These results are an important step towards low-latency real-time GW detection, enabling multi-messenger astronomy. 
\end{abstract}
\end{frontmatter}

\section{Introduction} 
\label{sec:intro}

The first gravitational wave (GW) detection on 2015 September 14 \cite{LIGOScientific:2016aoc} by the the advanced Laser Interferometer Gravitational-Wave Observatory (LIGO) and Virgo Collaboration \cite{LIGOScientific:2014pky,VIRGO:2014yos} ushered in a new era of GW Astrophysics. During the first (O1) and second (O2) observing runs the LIGO and Virgo collaborations reported eleven GW signals \cite{LIGOScientific:2018mvr} from compact binary mergers, which included ten binary black-hole (BBH) events and one clear binary neutron star (BNS) merger, GW170817 \cite{LIGOScientific:2017vwq}. The observation of GW170817 in both gravitational and electromagnetic (EM) radiation inaugurated the field of Multi-Messenger Astrophysics (MMA), which uses GWs, EM radiation, cosmic rays, and neutrinos to provide complimentary information about the astrophysical processes and environments of the sources \cite{LIGOScientific:2017vwq,LIGOScientific:2017ync}. The third observing run (O3) was split into two parts, O3a and O3b, and led to the detection of over 70 new GW events, including one additional BNS merger \cite{LIGOScientific:2020aai} and for the first time, two NSBH mergers \cite{LIGOScientific:2021qlt,LIGOScientific:2021djp}. Thus, including the candidates from O3, the recently released third Gravitational-Wave Transient Catalog (GWTC-3) contains over 90 events that include all configurations of compact object mergers \cite{LIGOScientific:2021djp}.

During the upcoming fourth observing run (O4), scheduled to commence in March 2023, and subsequent runs, the detection rates are expected to substantially increase with the greater instrument sensitivity \cite{KAGRA:2013rdx,Cahillane:2022pqm}. Moreover, with the advent of next-generation GW detectors in the next decade (Einstein Telescope and Cosmic Explorer \cite{LIGOScientific:2016wof,Maggiore:2019uih,Evans:2021gyd}), it is anticipated that millions of events will be detected per year, probing compact object mergers across the bulk of cosmic history. With the rapid increase in detection rate, the data analysis capabilities and techniques will have to grow and adapt.

At present, the detection of GW signals from compact binary mergers is primarily achieved using conventional matched-filtering methods that rely on large banks of pre-calculated waveform templates \cite{LIGOScientific:2021djp,Gabbard:2017lja,DalCanton:2014hxh}. Each template is a unique combination of a waveform model and source parameters, such as binary component masses and/or spins \cite{Bohe:2016gbl}. To generate a signal-to-noise ratio (SNR) time series, these methods take an inner product between the detector data and each template in the bank. However, the vast parameter space covered by the template bank due to the unknown source parameters, makes these approaches computationally challenging and expensive. As the rate of GW detections increases, unexpected events with unique physical properties will be observed more frequently in the future. Considering additional effects such as eccentricity, precession, and higher order models requires millions of waveform templates to cover the parameter space of all potential signals \cite{Harry:2016ijz,Harry:2017weg,2022arXiv220714645D}, making these methods computationally prohibitive. This is especially the case for GW events involving neutron stars, where the prompt follow-up of electromagnetic (EM) counterparts is crucial. As a result, there is a critical need for more efficient detection and classification algorithms that can overcome the limitations of conventional matched-filtering methods \cite{LIGOScientific:2021djp,Gabbard:2017lja,DalCanton:2014hxh}.

In recent times, there has been an upsurge in the application of Deep Learning (DL) approaches \cite{lecun2015deep} in various scientific and technical arenas, to expedite research that would otherwise be computationally demanding and to catalyze scientific discovery \cite{Deiana:2021niw}. With the aid of GPU computing, these techniques have shown exceptional performance in tasks like image recognition \cite{he2016deep} and natural language processing \cite{young2018recent}. Furthermore, DL has emerged as a new tool in engineering and scientific applications, supplementing traditional High Performance Computing (HPC), and has led to the evolution of a new field called Scientific Machine Learning \cite{baker2019workshop}. In recent years, there has been a growing interest in applying DL techniques in the field of GW astrophysics (see e.g., Refs. \cite{Huerta:2020xyq,Huerta:2021ybd,Cuoco:2020ogp}). Specifically, the use of Convolutional Neural Network (CNN) algorithms \cite{lecun1998gradient} has been pioneered, and has shown promising results in detecting simulated signals from BBH collisions embedded in Gaussian noise with performance that is comparable or even better than that of conventional matched-filtering methods \cite{Gabbard:2017lja,George:2016hay}. As a result, a growing number of research groups have begun to apply DL algorithms to detect GW BBH events, both in simulated Gaussian noise and realistic LIGO data (see e.g., Refs. \cite{George:2017pmj,Gebhard:2019ldz,Wang:2019zaj,Lin:2020aps,Morales:2020ksv,Xia:2020vem,Schafer:2020kor,Schafer:2021cml,Schafer:2021fea,Schafer:2022dxv}). These exciting developments demonstrate the potential of DL approaches to transform the field of GW astrophysics and to enhance our understanding of the Universe.

In our previous work we applied, for the first time, a DL approach to detect GW signals from BNS mergers, embedded in both simulated Gaussian noise \cite{Krastev:2019koe} and real LIGO data \cite{Krastev:2020skk}, and distinguish them from detector noise and BBH signals. In our later work \cite{Krastev:2020skk} we applied a CNN to successfully recover and classify all eleven GW events from the first public catalog, GWTC-1 \cite{LIGOScientific:2018mvr}.

In this article, we extend our detection/classification deep neural network to include the NSBH event category. This allows us to address the detection and classification of GW signals from all compact binary coalescence (CBC) configurations consistently in a unified DL framework. We show for the first time that a neural network can be used to detect GW signals from NSBH mergers embedded in highly non-stationary, non-Gaussian noise. Most importantly, we demonstrate that our DL approach is able to recover all \textit{real} GW events involving neutron stars to date -- the two BNS (GW170817 and GW190425) and the two NSBH (GW191219 and GW200115) mergers. These results are an important step towards real-time detection of gravitational waves from BNS and NSBH mergers, enabling prompt follow-up of EM counterparts of these important GW transients and multi-messenger astrophysics.

\section{Methods} 
\label{sec:methods}

Following Krastev \cite{Krastev:2019koe} and Krastev \textit{et al.} \cite{Krastev:2020skk}, we construct a large data-set of templates of real LIGO noise with injected simulated BBH, BNS, and NSBH waveform signals. Then, we train a CNN to discriminate between these three classes and noise and evaluate its performance on new, unseen injections, as well as on real GW events from GWTC-3 \cite{LIGOScientific:2021djp}.

\subsection{Dataset Construction} 
\label{sec:dataset}

We obtained real LIGO data from the LIGO and Virgo Gravitational Wave Open Science Center \cite{LIGOScientific:2019lzm}. Specifically, we used O2 and O3b data from the LIGO Livingston detector (L1) sampled at 4096 Hz which does not contain known GW events and hardware injections. 

To simulate GW CBC signals, we use the LIGO Algorithm Library Suite (LALSuite) \cite{2020ascl.soft12021L} to generate BBH, BNS, and NSBH waveforms. In particular, we used the \verb|SEOBNRv4| \cite{Bohe:2016gbl}, \verb|TaylorF2| \cite{Messina:2019uby}, and \verb|SEOBNRv4_ROM_NRTidalv2_NSBH|~\cite{Dietrich:2018uni} time domain approximants to generate BBH, BNS, and NSBH waveforms, respectively. For BBH waveforms, we uniformly sample component masses between $2$ and $95$ $M_\odot$ with a maximum mass ratio $m_1/m_2 \leq 10$. For BNS waveforms, we sample component masses uniformly between $1$ and $2$ $M_\odot$. Finally, for NSBH waveforms, we sample NS component masses uniformly between $1$ and $2$ $M_\odot$ and BH component masses uniformly between $2$ and $35$ $M_\odot$.

For all waveforms, we assume both components have zero spin and the binary system has zero eccentricity. For BNS and NSBH waveforms, we use the APR equation of state \cite{apr_eqn_state} to obtain the contribution from the tidal deformability parameter $\Lambda$ of the component neutron star(s); for calculating $\Lambda$, see, e.g., Refs. \cite{Hinderer:2009ca} and \cite{Krastev:2018nwr}. We sample waveforms at 4096 Hz for 4 seconds, which we have found is sufficient for achieving strong discrimination between each represented CBC class, and for recovering real GW signals from all CBC configurations. The shorter templates also help to reduce the memory requirements during the neural network training. Both the data and the simulated signals are whitened separately with power spectral density (PSD) computed directly from the raw GW strain data by Welch's method \cite{welch1967use}. Whitening of data is an operation of rescaling the noise contribution at each frequency to have equal power \cite{Gabbard:2017lja}. Because whitening is a linear procedure, whitening both parts individually is equivalent to whitening their sum. Subsequently, we position the peak of the waveform uniformly at random between 3.7 and 3.9 s in the template to make the training process robust to moderate time translations in the signal. We scale the injected waveform amplitude to achieve a particular signal-to-noise ratio (SNR), defined as \cite{Gabbard:2017lja}: 
\begin{equation}
    \rho^2_\mathrm{opt} = 4 \int_{f_\mathrm{min}}^\infty \frac{\left| \tilde{h}(f) \right|^2}{S_n(f)} df,
\end{equation}
where $\tilde{h}$ is the frequency domain representation of the GW strain, $S_n(f)$ is the single-sided detector noise PSD, and $f_\mathrm{min}$ is the frequency of the GW signal at the start of the sample time series. From an astrophysical perspective, rescaling the waveform is equivalent to varying the source distance from the detector. 
\begin{figure}[t]
  \centering
  \includegraphics[width=0.40\textwidth]{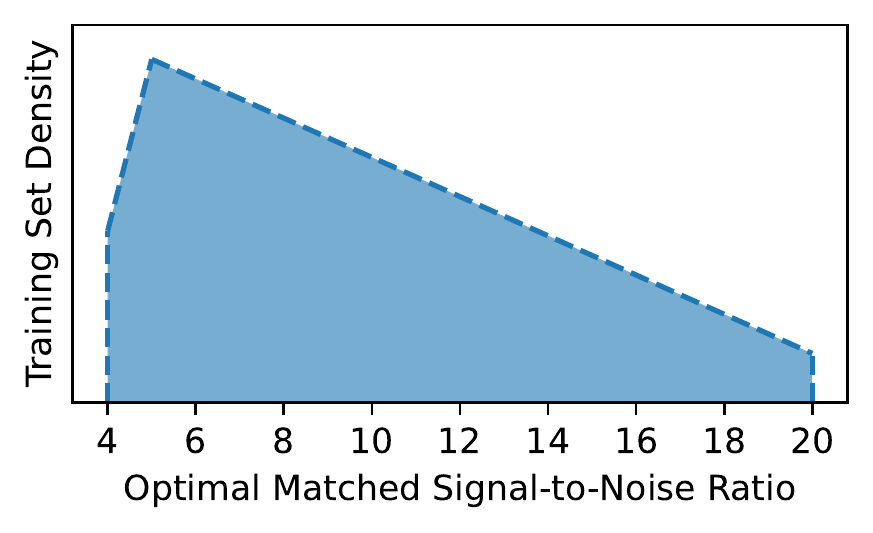}
  \caption{The distribution from which we sample training signal-to-noise ratios. We emphasize low SNR events in our training set to improve model generalization performance to other low SNR events.
  \label{fig:snr_dist}}
\end{figure}

We generate 480,000 templates for training, 16,000 templates for validation, and 1,600,000 templates for testing, all of which are disjoint. Each dataset is approximately $1/4$ noise with no event waveform, $1/4$ noise + BBH signal, $1/4$ noise + BNS signal, and $1/4$ noise + NSBH signal. For validation, we sample SNRs uniformly between 5 and 20. For testing, we sample SNRs uniformly between 1 and 20. And for training, to emphasize low SNR detections, we sample SNRs between 4 and 20 using a truncated triangular distribution with lower limit 3, mode 5, and upper limit 27.5 SNR; see \autoref{fig:snr_dist}. We find empirically that this distribution of SNRs for training improves our sensitivity, accuracy, and ability to recover confirmed events from the LIGO catalog. 

\subsection{Neural Network Architecture and Training}\label{sec:nn}

\begin{figure}[t]
  \centering
  \includegraphics[width=0.45\textwidth]{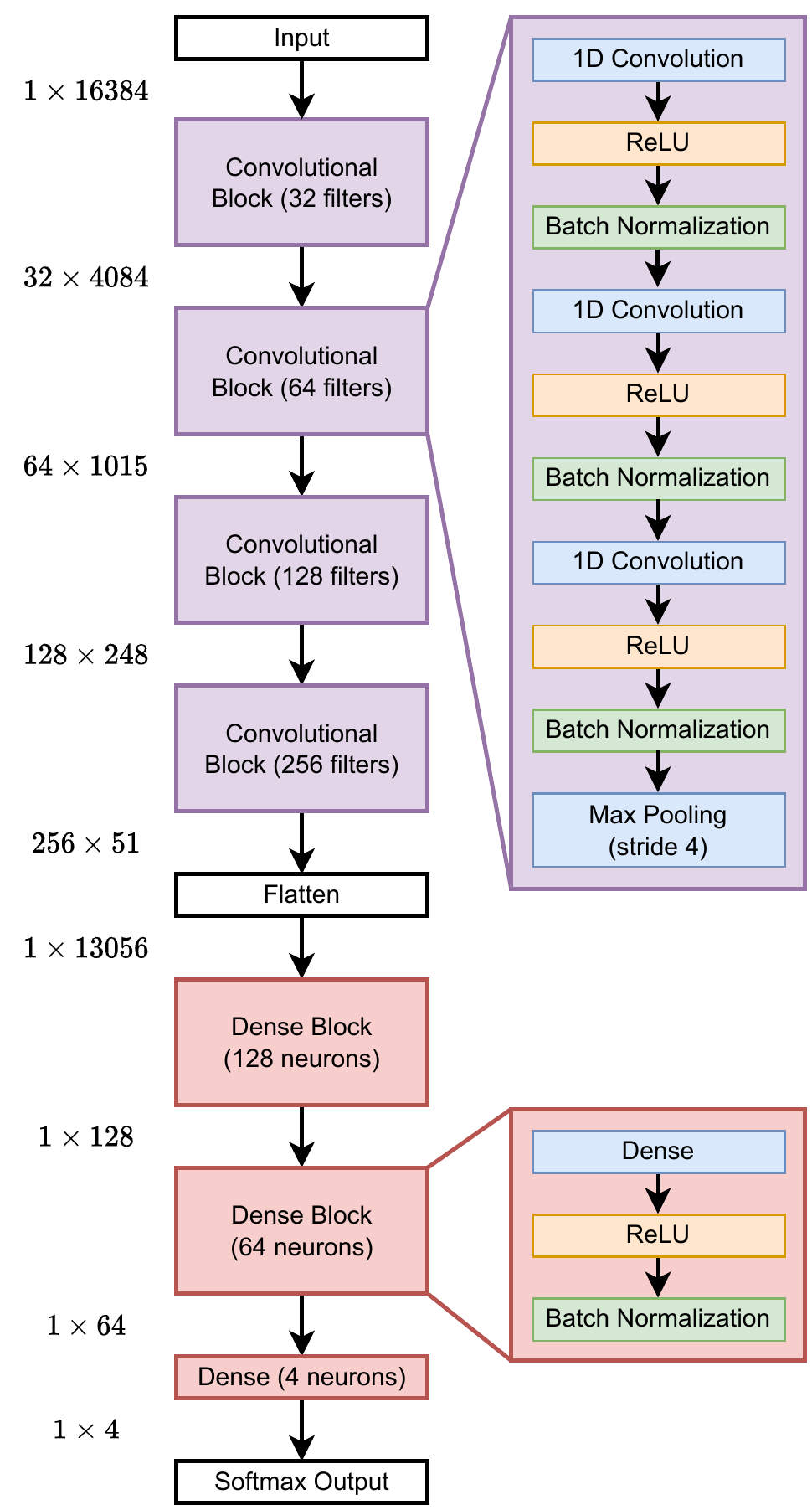}
  \caption{Schematic diagram of our neural network architecture. All 1D convolution layers within a given convolutional block have the same number of filters. Our model contains 3,441,380 parameters. 
  \label{fig:nn_architecture}}
\end{figure}

To choose a CNN architecture, we performed Bayesian hyperparameter optimization over architecture choices and learning rates. The final optimized architecture contained 4 convolutional blocks followed by 2 dense hidden layers and a softmax output. Each convolutional block contains 3 convolutional layers, each followed by a ReLU activation and a batch normalization layer and the entire block ends with a max pooling layer. The filter sizes within a given convolutional block are the same and chosen to be 16, 8, 8, and 8 for each respective convolutional block in the network. Each max pooling layer has filter size 4. The number of filters within a given convolutional block is also constant, and chosen to be 32, 64, 128, and 256. The dense hidden layers have widths 128 and 64. The final softmax output corresponds to the number of predicted classes, 4. The model has 3,441,380 parameters in total. A schematic diagram showing the architecture is provided in \autoref{fig:nn_architecture}.

We built and trained our CNN models using TensorFlow 2.9 \cite{tensorflow2015-whitepaper}. We performed hyperparameter optimization and experiment tracking with Weights and Biases \citep{wandb}. To train the models, we used the Adam \cite{kingma2014adam} optimizer with AMSgrad \cite{amsgrad}. Following hyperparameter optimization, we used $\alpha = 3.986 \times 10^{-3}$ as an initial learning rate, Adam $\beta_1 = 0.1888$, Adam $\beta_2 = 0.9537$, Adam $\epsilon = 1.4975 \times 10^{-3}$, and with a batch size of 256.  We used sparse categorical cross-entropy as a loss function and trained with a training budget of 50 epochs. Our final model is taken from the epoch with the lowest validation loss. We also use linear learning rate decay, decaying $\alpha$ by a factor of 0.1 over the first 30 epochs of training. We trained our model using 4 NVidia A100 GPUs with a data-parallel strategy. 

\section{Results} 
\label{sec:results}

We evaluate the performance of our model over a large testing dataset of synthetic GW event injections, described in \autoref{sec:dataset}. We also evaluate the performance of our model on real GW events using the GWTC-3 catalog \citep{LIGOScientific:2021djp}. This extended catalog includes the two BNS events (GW170817 and GW190425) and the two NSBH events (GW191219 and GW200115). Finally, we compare the performance of the trained DL model with models from our previous works \cite{Krastev:2019koe,Krastev:2020skk} that do not include the NSBH event class.

\subsection{Synthetic Event Detection}

\begin{figure}[t!]
  \centering
  \includegraphics[width=0.475\textwidth]{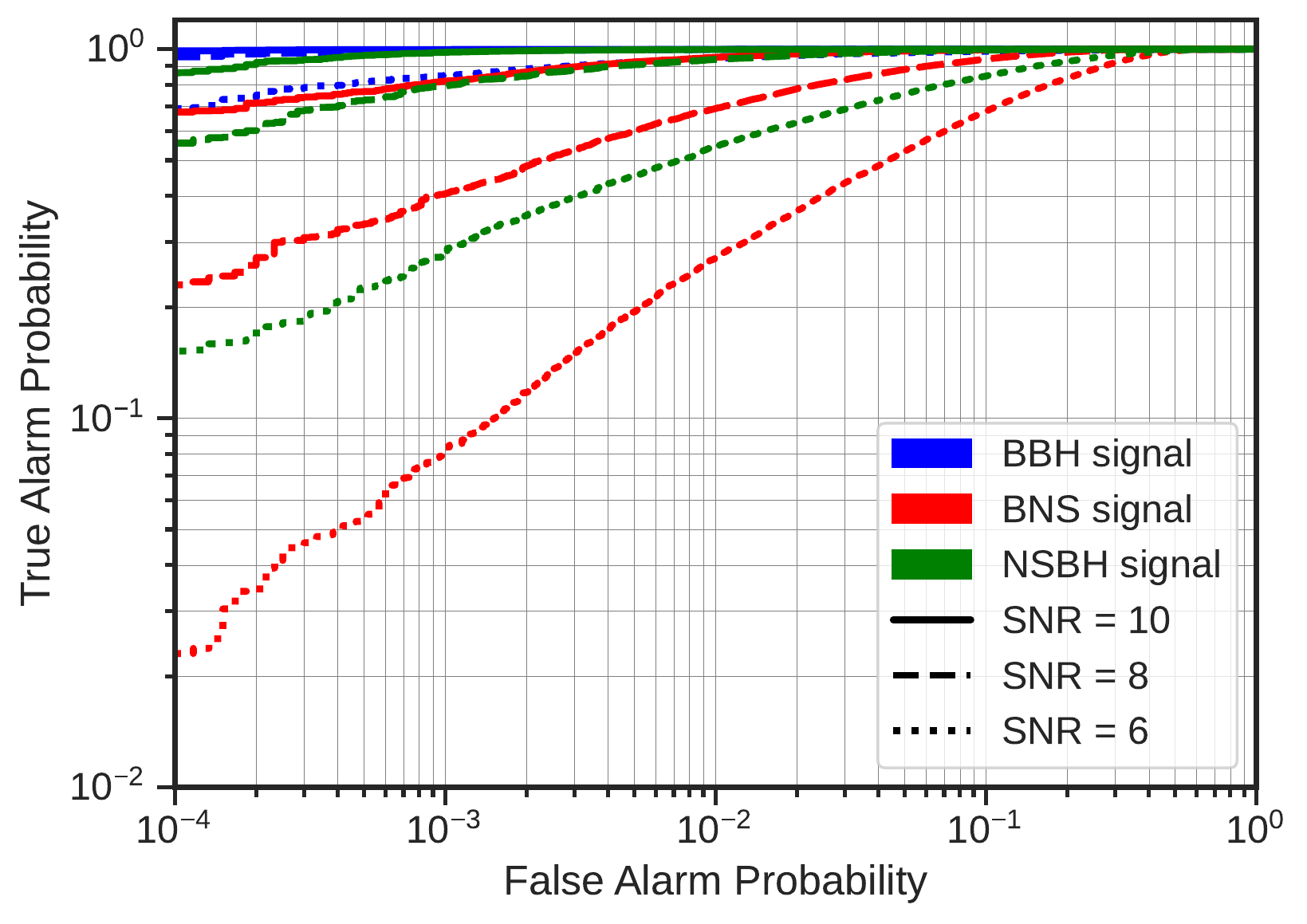}
  \caption{Receiver operating characteristic curves for all three CBC configurations at fixed signal-to-noise ratios, showing true positive rate and false positive rate, parameterized as a function of detection probability threshold. Note that axes are both shown in logarithmic scale. 
  \label{fig:roc}}
\end{figure}

\begin{figure}[t!]
  \centering
  \includegraphics[width=0.475\textwidth]{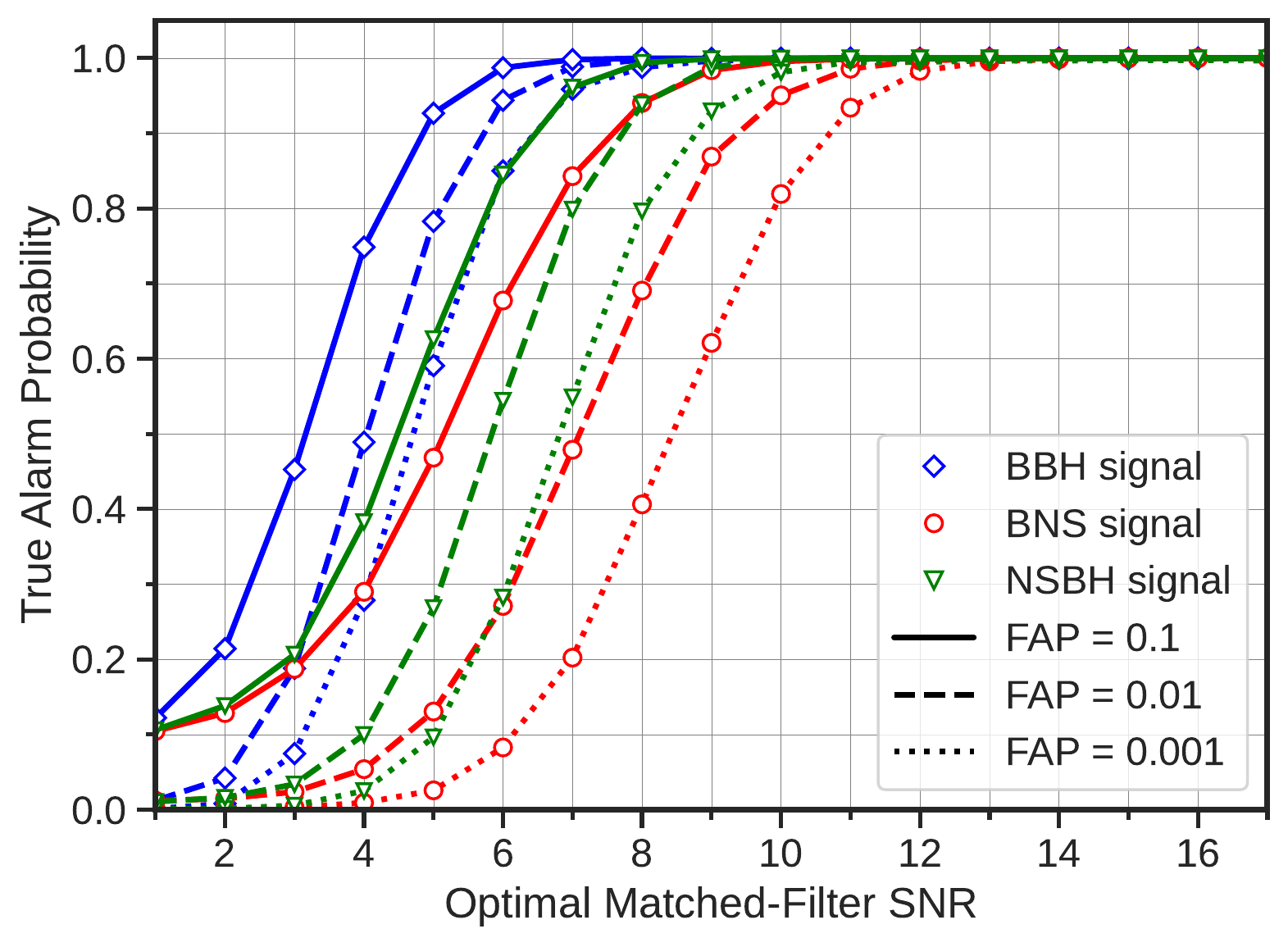}
  \caption{Sensitivity curves of the demonstrating the true positive rate at fixed signal-to-noise ratios and false positive rates for simulated BBH, BNS, and NSBH signals embedded in real LIGO noise. At $\rho_\mathrm{opt} \geq 13$, all signals are detected and correctly classified.
  \label{fig:sensitivity}}
\end{figure}

Following the same strategy as in our previous works~\cite{Krastev:2019koe,Krastev:2020skk}, we first consider the receiver operating characteristic (ROC) curves for each type of GW signal at a fixed SNR. To determine the ROC curve for a given type of CBC, we calculate the probability of each event in our test dataset being classified as that event type. A ROC curve then displays the proportion of true positives against the proportion of false positives parameterized as a function of the probability threshold to classify a given signal as an event. We calculate the ROC curves with the Python scikit-learn library (https://scikit-learn.org), which constructs empirical ROC curves. An empirical ROC curve shows the relationship between the true alarm probability (TAP) and the false alarm probability (FAP) for various threshold values. Each point on the curve corresponds to a different threshold value. In order to compare different ranking statistics, we can fix the FAP and choose the statistic with the highest TAP (sensitivity) at that FAP. This allows us to assess which statistic performs better overall. We varied the optimal SNR from 1 to 20 in integer steps of 1, and the trained model was applied to inputs with approximately equal fractions of each GW signal class (Noise, BBH Signal, BNS Signal, NSBH Signal).

We show the ROC curves calculated on our synthetic dataset in \autoref{fig:roc} for all three event classes at SNRs of 6, 8, and 10. These curves show that our model is most sensitive to BBH events, followed by NSBH events, and least sensitive to BNS events. We also find that our model is more sensitive to both BBH and BNS events than those presented in \cite{Krastev:2019koe, Krastev:2020skk}. Note that since the TAP is a function of the FAP, it also reaches a maximal sensitivity for signals with lower optimal SNR at a higher FAP.   

\begin{figure*}[t!]
  \centering
  \includegraphics[width=0.98\textwidth]{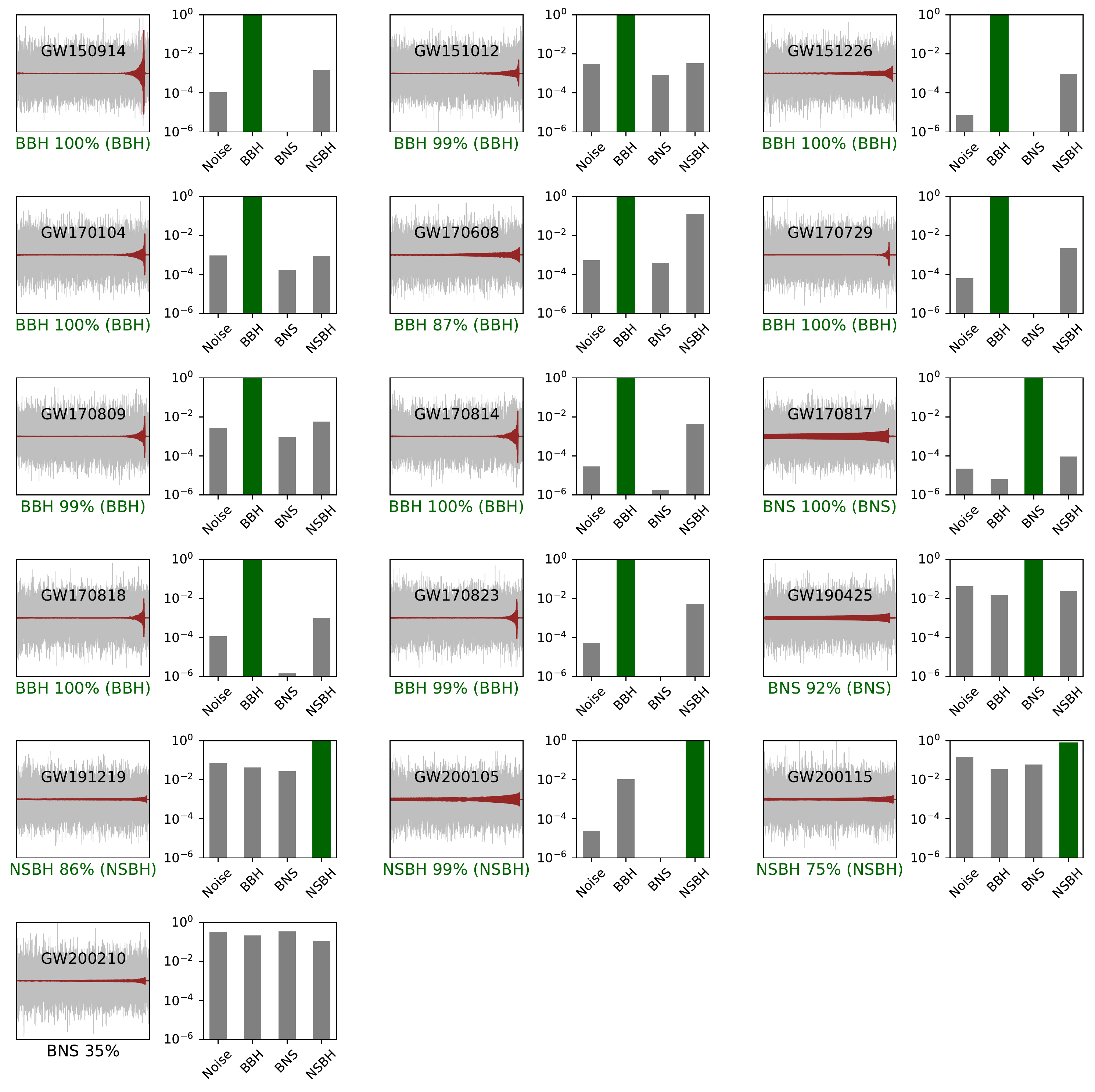}
  \caption{Detection of GWTC-1 events with our deep neural network. We also include BNS and NSBH (candidate) mergers from GWTC-2 and GWTC-3: GW170817 and GW190425 (BNS event from GWTC-2), GW191219 (NSBH event from GWTC-3), GW200105 (NSBH marginal candidate from GWTC-3), GW200115 (NSBH event from GWTC-3), and GW200210 (BBH or NSBH candidate event from GWTC-3). We show the probability of each event class predicted by our neural network for each GW event. The LIGO/Virgo inferred secondary mass for GW200210 is $2.83^{+0.47}_{-0.42}$, which is out of distribution for both our BNS and NSBH training datasets. Note that these probabilities are shown in logarithmic scale. 
  \label{fig:gwtc1_preds}}
\end{figure*}

We also examined the detection sensitivity of the classifier as a function of the SNR at a fixed FAP, shown in \autoref{fig:sensitivity}. Similar to ROC curves, we consider the fraction of true positives compared to false positives for all three CBC classes. However, we now plot the portion of true positives as a function of SNR at several representative FAPs ($0.1$, $0.01$, and $0.001$). These sensitivity curves represent the ability of the detection CNN to identify and classify GW signals from all three CBC event configurations (BBH, BNS, and NSBH). The lowest FAP used in our analysis translates to a false alarm rate (FAR) of 0.1\% or an estimated FAR of $\sim 10^3$ per month\footnote{To estimate the FAR, overlapping time-series segments of duration 0.2 s are used to match the length of the interval in which the signal peak varies. The FAR is calculated from the FAP, which is then converted to false alarms per month.}. The FAR can be decreased by applying the classifier independently to multiple GW detectors and enforcing coincidence \cite{Huerta:2020xyq,Wei:2020ztw,Chaturvedi:2022suc}, and also by checking the consistency of the estimated GW source parameters.  As before, we observe that our model is the most sensitive to BBH signals, followed by NSBH signals, and finally BNS signals. At a FAP of $0.001$, our model saturates the sensitivity curves for BBH events at $\rho_\mathrm{opt} \geq 8$, for NSBH events at $\rho_\mathrm{opt} \geq 10$, and for BNS events at $\rho_\mathrm{opt} \geq 13$. In summary, all curves saturate (at 1) for optimal ${\rm SNR} \geq 13$, which implies that all signals are always detected. Again, we note that, for BBH and BNS signals, our model is more sensitive than the models presented in our previous work \cite{Krastev:2019koe, Krastev:2020skk}. 

\subsection{Recovering Real GW Events}

\begin{table*}[t!]
\centering
\begin{tabular}{c|c|c|c|c}
\multirow{2}{*}{Event} & LIGO/Virgo Inferred & CNN Inferred & Maximum Single- & Detector \\
& Event Type & Event Type & Detector SNR & \\
\hline
GW190426\_152155 & BBH & Noise & --- & --- \\
GW190814 & BBH/NSBH & NSBH & --- & --- \\
GW190924\_021846 & BBH & NSBH & --- & --- \\
GW191103\_012549 & BBH & Noise & 6.9 & LIGO Livingston \\
GW191126\_115259 & BBH & Noise & 6.2 & LIGO Livingston \\
GW200208\_222617 & BBH & Noise & 6.0 & LIGO Livingston \\
GW200210\_092254 & BBH/NSBH & BNS & 8.0 & LIGO Livingston \\
GW200306\_093714 & BBH & Noise & 6.1 & LIGO Livingston \\ 
\hline
\end{tabular}
\caption{Events of interest or misclassified events in the GWTC-2 and GWTC-3 catalogs. Ref.~\cite{LIGOScientific:2020ibl} does not report individual detector SNRs and thus GWTC-2 events (GW190426\_152155, GW190814, and GW190924\_021846) are missing this data. We note that in follow-up analysis, GW190426\_152155 is reclassified as a marginal event \citep{LIGOScientific:2021usb}. LIGO/Virgo analysis finds that GW190814 is likely a BBH merger though a NSBH merger is also possible \citep{LIGOScientific:2020zkf}; interestingly, our pipeline classifies GW190814 as a NSBH event.} All misclassified BBH merger signals in GWTC-3 have single detector SNRs $\rho < 7.0$. Finally, the LIGO/Virgo inferred second component mass of $2.83^{+0.47}_{-0.42}$ $M_{\odot}$ for GW200210\_092254 is out-of-distribution for our BNS and NSBH training datasets, which have NS masses up to 2 $M{\odot}$. Although both binary component masses of this system are within the BBH mass range of our training datasets, the DL model cannot confidently classify it into any specific event class. See text for details.
\label{tab:misclassifications}
\end{table*}

To evaluate the applicability of our model beyond synthetic data, we apply our CNN to real GW strain data containing all events in the GWTC-1 \cite{LIGOScientific:2018mvr}, GWTC-2 \cite{LIGOScientific:2020ibl}, and GWTC-3 \cite{LIGOScientific:2021djp} catalogs, which include 82 BBH mergers, 2 BNS mergers, and 2 NSBH mergers. We obtained GW strain data for these events from the LIGO GWOSC through the catalogs provided by PyCBC \cite{pycbc}, and preprocessed them following the procedure described in \autoref{sec:dataset}. For events where multiple detectors were online, we selected the event data from the detector with the highest single-detector SNR.

In \autoref{fig:gwtc1_preds}, we shown results for all events from the GWTC-1 catalogue, as well as the two BNS mergers, the two confirmed NSBH merger signals, and two potential NSBH candidates. For brevity, in the Figure we omit results for the remaining BBH events from the GWTC-2 and GWTC-3 catalogs. The trained model recovers and classifies successfully all BNS and NSBH events, and 64 of the 72 BBH events (counting GW190814 and GW200210\_092254 as BBH) across both catalogs. Most importantly, we show for the first time that a deep neural network can identify \textit{real} GW signals from the NSBH CBC class. 

As shown in the Figure, our model successfully identifies all merger events involving neutron stars (BNS and NSBH) to date and distinguish them from BBH events and detector noise with very high confidence. This result is particularly important in the context of the next generation GW detectors where, due to the very high rate of detections, it will be critical to promptly identify events for EM follow-up. Moreover, the results summarized in \autoref{fig:gwtc1_preds} illustrate for the first time that GWs from all CBC classes can be detected and classified consistently in a unified DL based framework. 

Of the misclassified events, three are from the GWTC-2 catalog (O3a run) and five are from the GWTC-3 catalog (O3b run). These are summarized in \autoref{tab:misclassifications}. Of these events, we note that GW190426\_152155 is an exceptional event with a large FAR, and eventually demoted to a marginal candidate in a follow-up analysis \citep{LIGOScientific:2021usb}. GW190814 has an inferred secondary component mass of $\sim 2.6 M_\odot$ and it is possible that the merger is a NSBH rather than a BBH \citep{LIGOScientific:2020zkf}. It is noteworthy that our DL model identifies GW190814 as a NSBH event, despite both components falling within the BBH mass range of $[2-95] M_{\odot}$. The total mass of the GW190924\_021846 BBH merger is merely 13.9 $M_{\odot}$, making it the BBH coalescence with the lowest mass, and this particular attribute may have led to its erroneous classification as a NSBH merger. Moreover, during the occurrence of GW190924\_021846, a moderate glitch was observed, and it was among the signals marked for glitch removal \cite{Cornish:2021wxy}, which could have also caused its misclassification.

Of the misclassified events from O3b, we note that the inferred secondary component mass in the case of GW200210\_092254 is $2.83^{+0.47}_{-0.42}$ $M_{\odot}$, which puts the full 90\% credible interval outside the neutron star mass distributions of our training datasets, which have neutron star masses up to 2 $M{\odot}$. As shown in \autoref{fig:gwtc1_preds}, even though the masses of both binary components of GW200210 fall within the BBH mass range of our training datasets, our DL model cannot accurately classify this system into any specific event class -- it only misclassifies it as a BNS event with a confidence level of 35\%. Finally, we note that all remaining misclassified events from O3b are BBH merger signals with maximum (across all detection pipelines) single detector $\mathrm{SNR} < 7.0$, and as such are below the maximal sensitivity SNR range of our trained model for BBH signals (see \autoref{fig:sensitivity}). Thus, it is not surprising that these specific events were misclassified.

\subsection{Comparison with Other DL Models}

The DL algorithm presented in this work could be most readily compared with the methods developed in Refs.~\cite{Gabbard:2017lja,George:2016hay,George:2017pmj} as they are all employ 1-D CNNs to detect BBH GW signals embedded in both Gaussian and real LIGO noise. In particular, the sensitivity curves for the BBH signals shown in \autoref{fig:sensitivity} are very similar to the ones reported in these works. The major difference with the DL algorithms presented in these works is that our algorithm includes also the classification of BNS and NSBH GW signals embedded in real detector noise.

Very recently, Sch\"{a}fer \textit{et al.} \cite{Schafer:2022dxv} reported results from the first ML GW search mock data challenge (MLGWSC-1), including results from three algorithms based on the CNN architecture: MFCNN~\cite{Wang:2019zaj}, CNN-Conic~\cite{Gabbard:2017lja,Schafer:2021cml}, and TPI FSU Jena \cite{Schafer:2021fea}. The the CNN architecture employed by our DL algorithm compares most closely with the algorithms of CNN-Conic and TPI FSU Jena as they both are based on Ref. \cite{Gabbard:2017lja}. The major difference is that these algorithms are specifically trained to detect BBH GW signals, and subsequently applied to data streams from two detectors. On the other hand, our algorithm is designed and trained to detect and classify all CBC event types (BBH, BNS, and NSBH) in a consistent DL framework, but currently still operates on a single-detector data. In order to robustly compare the performance of our DL approach to that of the algorithms from Ref. \cite{Schafer:2022dxv}, our DL algorithm needs to be tested on the datasets from MLGWSC-1. However, this is beyond the scope of the current work, and it is left for following articles.

We also compared the performance of the current DL model with models from our previous works \cite{Krastev:2019koe,Krastev:2020skk}, which did not include the NSBH event class in the training datasets. Previous models that lacked the NSBH classification performed very poorly on test data containing injected NSBH signals where most signals were misclassified as detector noise, or BBH signals. Additionally, the models from our previous works failed on data containing \textit{real} NSBH events (GW191219, GW200105 and GW200115). The NSBH events were misclassified as BBH signals (GW191219 and GW200105) and noise (GW200115) respectively.

In comparison, our current DL model trained on a dataset containing the NSBH GW event type improves dramatically the performance on test data with injected simulated NSBH signals. As seen in \autoref{fig:sensitivity}, our model identifies 100\% of the injected NSBH signals with $SNR\geq13$ in our test dataset. Moreover, the DL model trained on data containing NSBH templates is able to successfully recover real NSBH events (see \autoref{fig:gwtc1_preds}). The presented results indicate a significant enhancement in the performance of the current model as compared to the previous models which did not include the NSBH GW event class. These results highlight the significance of the current study, particularly in the context of the precise identification of GW signals arising from CBC events involving neutron stars, where timely observation of the EM counterpart is of utmost importance.

\section{Discussion and Conclusion} \label{sec:conclusion}

We have demonstrated for the first time that a deep learning algorithm can detect and distinguish GW signals from NSBH mergers in real advanced LIGO data. We have shown that our model achieves high sensitivity on simulated injections of all three CBC classes. Critically, we have also applied the trained neural network to GW events in the GWTC-1, GWTC-2, and GWTC-3 catalogs, and we have shown that our model can recover all high-confidence BNS and NSBH events in these catalogs, as well as 74 of the 82 BBH events; the misclassified BBH signals are all of low SNR. These results are an important step towards a deep learning approach to real-time GW detection from multi-messenger sources, where rapid electromagnetic follow-up is critical. 

To further improve fidelity in realistic contexts, upcoming work will need to extend these deep learning algorithms to include realistic glitches and synthesize multiple detector data streams. The results in this work are also fundamental for real-time parameter estimation, where accurate and reliable classification is important. These DL based approaches can be extended to enable rapid parameter estimation, and employ Bayesian networks to quantify model uncertainty. 

\section*{Acknowledgements}

This research is supported in part by the NSF under grant AST-2108531 and Cooperative Agreement PHY-2019786 (The NSF AI Institute for Artificial Intelligence and Fundamental Interactions http://iafi.org). This work has made use of data, software and/or web tools obtained from the Gravitational Wave Open Science Center (https://www.gw-openscience.org), a service of LIGO Laboratory, the LIGO Scientific Collaboration and the Virgo Collaboration. LIGO is funded by the U.S. National Science Foundation. Virgo is funded by the French Centre National de Recherche Scientifique (CNRS), the Italian Istituto Nazionale della Fisica Nucleare (INFN) and the Dutch Nikhef, with contributions by Polish and Hungarian institutes. The computations in this paper were run on the FASRC Cannon cluster supported by the FAS Division of Science Research Computing Group at Harvard University.

\bibliographystyle{elsarticle-num}
\bibliography{refs}

\end{document}